\documentclass[showpacs,showkeys,preprintnumbers,amsmath,amssymb,latexsym]{revtex4}

\usepackage{graphicx}
\usepackage{dcolumn}
\usepackage{bm}

\begin{document}

\title{Charged massive particle at rest in the field of a Reissner-Nordstr\"om  black hole\\
II. Analysis of the field lines and the electric Meissner effect}

\author{D. Bini}
  \email{binid@icra.it}
 \affiliation{Istituto per le Applicazioni del Calcolo ``M. Picone,'' CNR I-00161 Rome, Italy and\\ 
ICRA, University of Rome ``La Sapienza,'' I-00185 Rome, Italy}

\author{A. Geralico}
  \email{geralico@icra.it}
 \affiliation{Physics Department and
ICRA, University of Rome ``La Sapienza,'' I-00185 Rome, Italy}

\author{R. Ruffini}
\email{ruffini@icra.it}
\affiliation{Physics Department and
ICRA, University of Rome ``La Sapienza,'' I-00185 Rome, Italy\\
and ICRANet, I-65100 Pescara, Italy and Universit\'e de Nice Sophia Antipolis, Grand Ch$\hat a$teau, BP 2135, 28, avenue de Valrose, 06103 NICE CEDEX 2, France}

\begin{abstract}
The properties of the electric field of a two-body system consisting of a Reissner-Nordstr\"om black hole and a charged massive particle at rest have recently been analyzed in the framework of first order perturbation theory following the standard approach of Regge, Wheeler and Zerilli. 
In the present paper we complete this analysis by numerically constructing and discussing the lines of force of the \lq\lq effective'' electric field of the sole particle with the subtraction of the dominant contribution of the black hole. We also give the total field due to the black hole and the particle. As the black hole becomes extreme an effect analogous to the Meissner effect arises for the electric field, with the \lq\lq effective field'' lines of the point charge being expelled by the outer horizon of the black hole. This effect existing at the level of test field approximation, i.e. by neglecting the backreaction on the background metric and electromagnetic field due to the particle's mass and charge, is here found also at the complete perturbative level.  
We point out analogies with similar considerations for magnetic fields by Bi{\v c}\'ak and Dvo{\v r}\'ak.  
We also explicitly show that the linearization of the recently obtained Belinski-Alekseev exact solution coincides with our solution in the Regge-Wheeler gauge.
Our solution thus represents a \lq\lq bridge'' between the test field solution, which neglects all the feedback terms, and the exact two-body solution, which takes into account all the non-linearity of the interaction.
\end{abstract}



\maketitle

\section{Introduction}

We recently presented a perturbative solution describing a two-body system consisting of a Reissner-Nordstr\"om black hole with mass $\mathcal{M}$ and charge $Q$ and a charged massive particle with mass $m$ and charge $q$ at rest \cite{bgr_pla,bgr_prd}.
The system of the Einstein-Maxwell equations was solved by using the first order perturbation approach formulated by Zerilli \cite{Zerilli} based on the tensor harmonic expansion of both the gravitational and electromagnetic fields and using the Regge-Wheeler gauge condition \cite{ReggeW}. 
Closed form expressions for both the perturbed metric and electromagnetic field have been explicitly given, including the contribution of both the \lq\lq electromagnetically induced gravitational perturbation'' and \lq\lq gravitationally induced electromagnetic perturbation'' \cite{jrz2,jrz1}.  

The results discussed in \cite{bgr_pla,bgr_prd} gave answer to a problem whose investigation started long ago by Hanni and Ruffini \cite{HR}. They obtained the solution for a charged particle at rest in the field of a Schwarzschild
black hole in the case of test field approximation, i.e. under the conditions $q/m\gg1$, $m\approx0$ and $q\ll{\mathcal M}$, $q\ll Q$, by using the vector harmonic expansion of the electromagnetic field in curved space. The conditions above imply the solution of the Maxwell equations only in a fixed Schwarzschild metric, since the perturbation to the background geometry given by the electromagnetic stress-energy tensor is second order in the particle's charge and the effect of the particle's mass is there neglected. As a result, no constraint on the position of the test particle follows from the Einstein equations and the Bianchi identities: the position of the particle is totally arbitrary. 

This same test field approximation has been applied to the case of a Reissner-Nordstr\"om black hole by Leaute and Linet \cite{leaute}. In analogy with the Schwarzschild case, they used the vector harmonic expansion of the electromagnetic field holding the background geometry fixed. However, this \lq\lq test field approximation'' is not valid in the present context. In fact, in addition to neglecting the effect of the particle mass on the background geometry, this treatment also neglects the electromagnetically induced gravitational perturbation terms linear in the charge of the particle which would contribute to modifying the metric as well. 

The inclusion of the mass in such a simplified analysis, neglecting the feedback terms, is due to Bonnor \cite{bonnor}. He studied the condition for the equilibrium involving the black hole and particle parameters $Q, {\mathcal M}, q, m$ as well as their separation distance $b$.
He found the constraint
\begin{equation}
\label{bonnoreqcond0}
m=\frac{qQb}{{\mathcal M}b-Q^2}\left(1-\frac{2{\mathcal M}}{b}+\frac{Q^2}{b^2}\right)^{1/2}\ .
\end{equation}  
It follows that equilibrium exists independent of the separation if and only if the black hole is extreme, i.e. $|Q|/\mathcal{M}=1$, and the particle has the same ratio $|q|/m=1$.
In the general non-extreme case $|Q|/\mathcal{M}<1$ there is instead only one position of the particle which corresponds to equilibrium, for given values of the charge-to-mass ratios of the bodies. In this case the particle charge-to-mass ratio satisfies $|q|/m>1$, corresponding to a naked singularity.

The simplest special case of a massive neutral particle at rest near a Schwarzschild black hole has been studied in \cite{bgr_prd} in the framework of first order perturbation theory. We first showed explicitly that a perturbative solution for this problem free of singularities cannot exist. We then gave the explicit form of the perturbation  corresponding to a stable configuration when there is the presence of a \lq\lq strut'' between the particle and the black hole, corresponding to a conical singularity. 

We then solved in \cite{bgr_prd} the general case of a charged massive particle at rest in a Reissner-Nordstr\"om background by using both the vectorial and tensorial perturbations to describe the electromagnetic and gravitational perturbed fields respectively.
The perturbed metric we derived is spatially conformally flat, free of singularities. The perturbed electromagnetic field satisfies the Gauss' theorem, a fact that guarantees the correctness of the adopted boundary conditions.
The equilibrium condition for the system turned out to be the same as the condition (\ref{bonnoreqcond0}) obtained by Bonnor in his simplified approach. 
This is surprising, since our result has been obtained within a more general framework, and both the gravitational and electromagnetic  fields are different from those used by Bonnor.

We analyze in the present paper the properties of the perturbed electric field with special attention to the construction of the lines of force of the electric field. The two cases have been considered of the sole particle, with the subtraction of the dominant contribution of the black hole, as well as of the total field due to the black hole and the particle.
As the black hole becomes extreme an effect similar to the ordinary Meissner effect for magnetic fields in the presence of superconductors arises: the electric field lines of the point charge are expelled outside the outer horizon. 

The analogous phenomenon of expulsion of a magnetic field from extreme charged black holes was studied by Bi{\v c}\'ak and Dvo{\v r}\'ak \cite{bicdev2} in the same framework of first order perturbation theory, still adopting the Regge-Wheeler gauge. They used the hamiltonian formalism developed by Moncrief \cite{moncrief1,moncrief2,moncrief3} instead of the Zerilli's approach.
The equivalence between these two different treatments of perturbations have been extensively investigated by Bi{\v c}\'ak \cite{bicak}. 
They constructed the axially symmetric magnetic field of a current loop in the equatorial plane of an extreme Reissner-Nordstr\"om black hole as well as of a magnetic dipole placed on the polar axis, and the electromagnetic and gravitational fields occurring when a general non-extreme Reissner-Nordstr\"om black hole is placed in an asymptotically uniform magnetic field, finding in all cases that no line of force crosses the horizon as the black hole approaches the extreme condition.

The analogies between the phenomenon of expulsion of a magnetic field from extreme charged black holes and the ordinary magnetic Meissner effect was analyzed by Bi{\v c}\'ak and Ledvinka \cite{bicledv} also in the context of superconducting branes and extremal black holes in string theory, starting from developments of the evidence for the Meissner effect for extremal black hole solutions in string theory and Kaluza-Klein theory by Chamblin, Emparan and Gibbons \cite{emparan}.
It is appropriate to remark that in contrast to the magnetic Meissner effect considered by Bi{\v c}\'ak and Dvo{\v r}\'ak, the effective \lq\lq electric Meissner effect'' considered in the present article has no classical analogue, as far as we know, and is a pure general relativistic effect. 

Recently an important progress has been achieved by Belinski and Alekseev \cite{belinski}. They have obtained an exact two-body solution of the Einstein-Maxwell equations in explicit analytic form for the system consisting of a Reissner-Nordstr\"om black hole and a naked singularity, by using the monodromy transform approach \cite{alek}. Technical details on the construction of the solution have been given in \cite{belalekgrqc}. 
They have shown that an equilibrium without intervening struts or tensions is possible for such a system at selected values of the separating distance between the sources. 

More recently Manko \cite{manko2} have recalled the existence in the literature of an exact electrostatic multi-soliton solution to the Einstein-Maxwell equations corresponding to a Reissner-Nordstr\"om black hole in equilibrium with a naked singularity \cite{manko}, obtained by using the Sibgatullin's integral equation method \cite{sibgatullin}.
In contrast to the Belinski-Alekseev work, where an explicit analytic condition for the equilibrium was obtained, the equilibrium configurations were discussed in \cite{manko} only numerically. In fact, many difficulties were encountered there in relating the set of parameters of the solution to the physical parameters, i.e. the physical masses and charges as well as the separation distance between the bodies. 
Manko has indeed obtained in \cite{manko2} out of this formalism the analytic expression of the Belinski-Alekseev solution by very long algebraic manipulations. 

We show in the Appendix that the Belinski-Alekseev solution, once linearized with respect to the mass and charge of the naked singularity, coincides with our solution.
In this limit also the equilibrium condition obtained by Belinski and Alekseev exactly reduces to our equilibrium condition. 
Our first order result therefore confirms the validity and offers a tool for the physical interpretation of the Belinski-Alekseev solution. 

In Section II we review the properties of our solution derived in \cite{bgr_pla,bgr_prd}. 
In Section III we provide a definition of the electric field lines following \cite{HR}, drawing them both in the case of the \lq\lq effective'' electric field of the sole particle and of the total field due to the hole and particle. 
We then apply the concept of induced charge density on the horizon introduced in \cite{HR}, discussing how to overcome the problems related to its definition due to the difficulty of having well behaved horizon fields.
The embedding diagram corresponding to the effective field of the particle is also shown, allowing to visualize the effect of the spacetime curvature. 
The \lq\lq electric Meissner effect'' arising as the black hole becomes extreme is discussed in Section IV, and we finally proceed to the general conclusions.

\section{The analytic solution for the perturbed metric and electromagnetic field}

Let us briefly summarize the results and the properties of the solution derived in \cite{bgr_pla,bgr_prd}.
In standard Schwarzschild-like coordinates the Reissner-Nordstr\"om black hole metric is
\begin{eqnarray}
\label{RNmetric}
ds^2&=&- f(r)dt^2 + f(r)^{-1}dr^2+r^2(d\theta^2 +\sin ^2\theta d\phi^2)\ ,
\nonumber\\
f(r)&=&1 - \frac {2\mathcal{M}}{r}+\frac{Q^2}{r^2}\ ,
\end{eqnarray}
with associated  electromagnetic field
\begin{equation}
\label{RNemfield}
F=-\frac{Q}{r^2}dt\wedge dr\ .
\end{equation}
The horizons are located at $r_\pm={\mathcal M}\pm\sqrt{{\mathcal M}^2-Q^2}\equiv{\mathcal M}\pm\Gamma$; we consider the case $ |Q|\leq {\mathcal M} $ and the region $r>r_+$ outside the outer horizon, with an extremely charged hole corresponding to $|Q|={\mathcal M}$ (which implies $\Gamma=0$) where the two horizons coalesce.

We have considered in \cite{bgr_pla,bgr_prd} a perturbation of the Reissner-Nordstr\"om solution due to a point charge of mass $m$ and charge $q$ at rest at the point $r=b $ on the polar axis $\theta=0$. 
We will denote by a tilde the quantities which refer to the total electromagnetic and gravitational fields, to first order of the perturbation
\begin{eqnarray}
\label{pertrelations}
\tilde g_{\mu \nu }=g_{\mu \nu } + h_{\mu \nu }\ ,\qquad
\tilde F_{\mu \nu }=F_{\mu \nu }+  f_{\mu \nu }\ ,
\end{eqnarray}
while the corresponding quantities without the tilde refer to the background Reissner-Nordstr\"om geometry (\ref{RNmetric}) and electromagnetic field (\ref{RNemfield}).
It is worth noting that the perturbation will be small if $m$ and so $q$ are sufficiently small with respect to the black hole mass and charge whereas the charge to mass ratio need not to be small. 
The metric describing such a two-body system to first order of the perturbation is given by \cite{bgr_pla,bgr_prd}
\begin{eqnarray}
\label{lineelemnonextr}
d{\tilde s}^2=-[1-{\bar {\mathcal H}}]f(r)dt^2 + [1+{\bar {\mathcal H}}][f(r)^{-1}dr^2+r^2(d\theta^2 +\sin ^2\theta d\phi^2)]\ ,
\end{eqnarray}
where 
\begin{eqnarray}
\label{barHdef}
{\bar {\mathcal H}}&=&2\frac{m}{br}f(b)^{-1/2}\frac{(r-{\mathcal M})(b-{\mathcal M}) -\Gamma^2\cos\theta}{{\bar {\mathcal D}}}
=-2\frac{m}{br}f(b)^{-1/2}\frac{\partial {\bar {\mathcal D}}}{\partial\cos\theta}\ , \nonumber\\
{\bar {\mathcal D}} &=& [(r-{\mathcal M})^2+(b-{\mathcal M})^2 - 2(r -{\mathcal M})(b-{\mathcal M})\cos\theta- \Gamma^2\sin^2\theta]^{1/2}\ , 
\end{eqnarray}
and the condition $|{\bar {\mathcal H}}|\ll1$ must be satisfied according to the validity of the perturbation approach. 

The condition of having no singularities on the symmetry axis leads to the following equilibrium condition for the system 
\begin{equation}
\label{bonnoreqcond}
m=qQ\frac{b f(b)^{1/2}}{{\mathcal M}b-Q^2}\ ,
\end{equation}  
involving the black hole and particle parameters as well as their separation distance $b$.
If the black hole is extreme (i.e. $|Q|/\mathcal{M}=1$), then the particle must also have the same ratio $|q|/m=1$, and equilibrium exists independent of the separation; the solution thus reduces to the linearized form of the well known exact solution by Majumdar and Papapetrou \cite{maj,pap} for two extreme Reissner-Nordstr\"om black holes.
In the general non-extreme case $|Q|/\mathcal{M}<1$ there is instead one and only one position of the particle which corresponds to the equilibrium (\ref{bonnoreqcond}), for given values of the charge-to-mass ratios of the bodies. In this case the particle charge-to-mass ratio must satisfy the condition $|q|/m>1$. 

The total perturbed electromagnetic field in its original form (see Eqs. (224)--(225) of \cite{bgr_prd}) turned out to be
\begin{equation}
\label{oldFmunu}
\tilde F=\left[-\frac{Q(1+{\bar {\mathcal H}}_0/2)}{r^2}+f_{tr}\right]dt\wedge dr +f_{t\theta}dt\wedge d\theta\ ,
\end{equation}
where ${\bar {\mathcal H}}_0=-2q\Gamma^2/[Q({\mathcal M}b-Q^2)]$,  
\begin{eqnarray} 
\label{Zeremtensorpertnonextr} 
f_{tr}=\partial_r V\ , \qquad  
f_{t\theta}=\partial_\theta V\ , 
\end{eqnarray} 
and $V$ denotes the first order electrostatic potential  
\begin{equation}
\label{Vpert}
V=\frac1r\left[\frac{{\mathcal M}r-Q^2}{2Q}{\bar {\mathcal H}}-\frac{Q{\bar {\mathcal H}}_0}{2r}\right]\ .
\end{equation}

As already noted in \cite{bgr_prd} this solution apparently does not satisfy Gauss' theorem (see Eq. (232) of \cite{bgr_prd})
\begin{equation}
\label{flussonew}
\Phi=\int_{S}{}^*{\tilde F}\wedge dS=4\pi[Q(1+{\bar {\mathcal H}}_0/2)+q\vartheta(r-b)]\ ,
\end{equation}
where $\Phi$ is the flux of the electric field obtained by integrating the dual of the electromagnetic form (\ref{oldFmunu}) over a spherical $2$-surface $S$ centered at the origin where the black hole charge $Q$ is placed and with variable radius ($r$ greater or lesser than $b$), the function $\vartheta(x)$ denoting the step function.
In order to get the correct result, in \cite{bgr_prd} (see Eq. (236) there) we have added to the electrostatic potential the extra term $\bar V=-Q{\bar {\mathcal H}}_0/(2r)$, which is a solution of the $l=0$ mode homogeneous Maxwell equations.
But due to the coupling of the system of Einstein-Maxwell equations the latter term gave back a modification of the metric function ${\bar {\mathcal H}}$ by the additional constant term ${\bar {\mathcal H}}_0$. 
In \cite{bgr_prd} we (only) argued the existence of a suitable gauge transformation to remove the constant term ${\bar {\mathcal H}}_0$ in this metric, keeping at the same time the Gauss' theorem be satisfied. 
Actually a direct inspection of Eqs. (\ref{oldFmunu})  and (\ref{flussonew}) shows that the most natural and simplest way to proceed is to replace (formally) the black hole charge parameter according to $Q\to Q(1-{\bar {\mathcal H}}_0/2)$. 
As a result, substituting this relation into both solutions for the metric and electromagnetic field leads to
\begin{eqnarray}
\label{belalekmetlin}
d{\tilde s}^2&=&-[1-{\bar {\mathcal H}}-k(r)]f(r)dt^2 + [1+{\bar {\mathcal H}}+k(r)]f(r)^{-1}dr^2+(1+{\bar {\mathcal H}})r^2(d\theta^2 +\sin ^2\theta d\phi^2)\ , \nonumber\\
k(r)&=&\frac{{\bar {\mathcal H}}_0 Q^2}{r^2f(r)}\ , 
\end{eqnarray}
and
\begin{equation}
\label{RNemfieldpertnonextr}
\tilde F=\left[-\frac{Q}{r^2}+f_{tr}\right]dt\wedge dr +f_{t\theta}dt\wedge d\theta\ ,
\end{equation}
which identically satisfy Gauss' theorem $\Phi=4\pi[Q+q\vartheta(r-b)]$.

Finally we note that the total perturbed electrostatic potential 
\begin{equation}
\label{Vtot}
V_{\rm tot}=\frac1r\left[\frac{{\mathcal M}r-Q^2}{2Q}{\bar {\mathcal H}}-\frac{Q{\bar {\mathcal H}}_0}{2r}+Q\right]\ 
\end{equation}
obtained by summing the first order potential (\ref{Vpert}) and the own contribution $V_{\rm{BH}}=Q/r$ of the black hole itself can be conveniently written as the sum of the electrostatic potential $V_{\rm test}$ of the particle alone obtained by Leaute and Linet \cite{leaute} within the test field approximation, plus the black hole contribution $V_{\rm{BH}}$, plus some interaction terms $V_{\rm{int}}$ representing the \lq\lq gravitationally induced'' as well as \lq\lq electromagnetically induced'' electrostatic potential
\begin{equation}
\label{relperttest}
V_{\rm tot}=V_{\rm test}+V_{\rm BH}+V_{\rm{int}}\ , 
\end{equation}
where 
\begin{eqnarray}
\label{solRNpot}
V_{\rm{test}} &=& \frac q{b r} \frac{(r-{\mathcal M})(b-{\mathcal M})
 -\Gamma^2\cos\theta}{{\bar {\mathcal D}}} + \frac{q{\mathcal M}}{b r}\ , \nonumber\\
V_{\rm{int}} &=&-\left[\frac12\left(1-\frac{r}b\right){\bar {\mathcal H}}
+\frac{qQ}{{\mathcal M}b-Q^2}\left(1-\frac{{\mathcal M}}b\right)\right]\frac{Q}{r}\ .
\end{eqnarray}

\section{Electric field lines and the induced charge on the horizon}

The electric field associated with an observer with four-velocity $U$ is given by 
\begin{equation}
\label{pertelecompts}
E(U)^\alpha=\tilde F^\alpha{}_\beta U^\beta\ .
\end{equation}
Correspondingly, the electric field lines are defined as the integral curves of the differential equation 
\begin{equation}
\label{ele1}
\frac{dx^\alpha}{d\lambda}=E(U)^\alpha\ ,
\end{equation}
where $\lambda$ is an affine parameter for the curves.
We select the static observers with respect to the metric (\ref{belalekmetlin}), whose four-velocity is given by
\begin{equation}
\label{staticobs}
U=\frac1{\sqrt{-\tilde g_{tt}}}\partial_t=f(r)^{-1/2}\left(1+\frac{{\bar {\mathcal H}}+k(r)}{2}\right)\partial_t\ ,
\end{equation}
to first order of the perturbation. Eq. (\ref{ele1}) thus becomes
\begin{equation}
\frac{dr}{d\lambda}=E(U)^r\ , \qquad 
\frac{d\theta}{d\lambda}=E(U)^\theta\ , 
\end{equation}
leading to the equation
\begin{equation}
\label{elelines4}
-E(U)^r\, d\theta+E(U)^\theta\, dr=0\ .
\end{equation}

From the Gauss' theorem we have that integrating the dual of the electromagnetic form (\ref{RNemfieldpertnonextr}) over a spherical $2$-surface $S$ centered about the black hole gives
\begin{equation}
\label{gausspert}
\Phi=\int_{S}{}^*{\tilde F}\wedge dS=4\pi[Q+q\vartheta(r-b)]\equiv\Phi^{(0)}+\Phi^{(1)}\ ,
\end{equation}
where the superscripts $(0)$, $(1)$ refer to the zeroth order and first order terms respectively.
In this case  the only nonvanishing components of ${}^*{\tilde F}$ are
\begin{eqnarray}
\label{fstarcompts}
{}^*{\tilde F}_{\theta\phi}&=&-r^2\sin\theta\left[-(1+{\bar {\mathcal H}})\frac{Q}{r^2}+f_{tr}\right]\equiv{}^*{\tilde F}_{\theta\phi}^{(0)}+{}^*{\tilde F}_{\theta\phi}^{(1)}\ , \nonumber\\
{}^*{\tilde F}_{r\phi}&=&f(r)^{-1}\sin\theta f_{t\theta}\equiv{}^*{\tilde F}_{r\phi}^{(1)}\ ,
\end{eqnarray}
so that the flux across a generic $2$-surface $S$ is given by:
\begin{equation}
\label{gausspert1}
\Phi=\int_{S} \,\left[{}^*{\tilde F}_{r\phi}dr\, d\phi+{}^*{\tilde F}_{\theta\phi}d\theta\, d\phi\right]\ .
\end{equation}
Therefore, as the electromagnetic field components do not depend explicitly on $\phi$, if $S$ is a generic revolution surface around the symmetry $z$-axis
we can write
\begin{equation}
\label{gausspert2}
\Phi=2\pi \int_{S} \,\left[{}^*{\tilde F}_{r\phi}dr +{}^*{\tilde F}_{\theta\phi}d\theta\right]\ ,
\end{equation}
so that the elementary flux across an infinitesimal closed surface, limited by the two spherical caps $\phi\in[0,2\pi]$, $\theta=\theta_0$ and $r=r_0$ and $\phi\in[0,2\pi]$, $\theta=\theta_0+d\theta$ and $r=r_0+dr$), is given by
\begin{equation}
\label{gausspert3}
d\Phi=2\pi [{}^*{\tilde F}_{r\phi}dr +{}^*{\tilde F}_{\theta\phi}d\theta]\ .
\end{equation}
The lines of constant electric flux ($d\Phi=0$) are then defined as those curves solutions of the equation
\begin{equation}
\label{elelines2}
0={}^*{\tilde F}_{r\phi}dr+{}^*{\tilde F}_{\theta\phi}d\theta\ .
\end{equation}
The important property is that for a static spacetime and using a static family of observers the constant flux lines coincide with the 
electric lines of force \cite{bgj}. In fact, we have
\begin{equation}
\label{elelines3}
{}^*{\tilde F}_{\theta\phi}= -\frac{\sqrt{-\tilde g}}{U_0}E(U)^r\ , \qquad 
{}^*{\tilde F}_{r\phi}=  \frac{\sqrt{-\tilde g}}{U_0}E(U)^\theta\ ,
\end{equation}
and Eq. (\ref{elelines2}) reduces exactly to the form (\ref{elelines4}).

\begin{figure} 
\typeout{*** EPS figure 1}
\begin{center}
$\begin{array}{cc}
\includegraphics[scale=0.45]{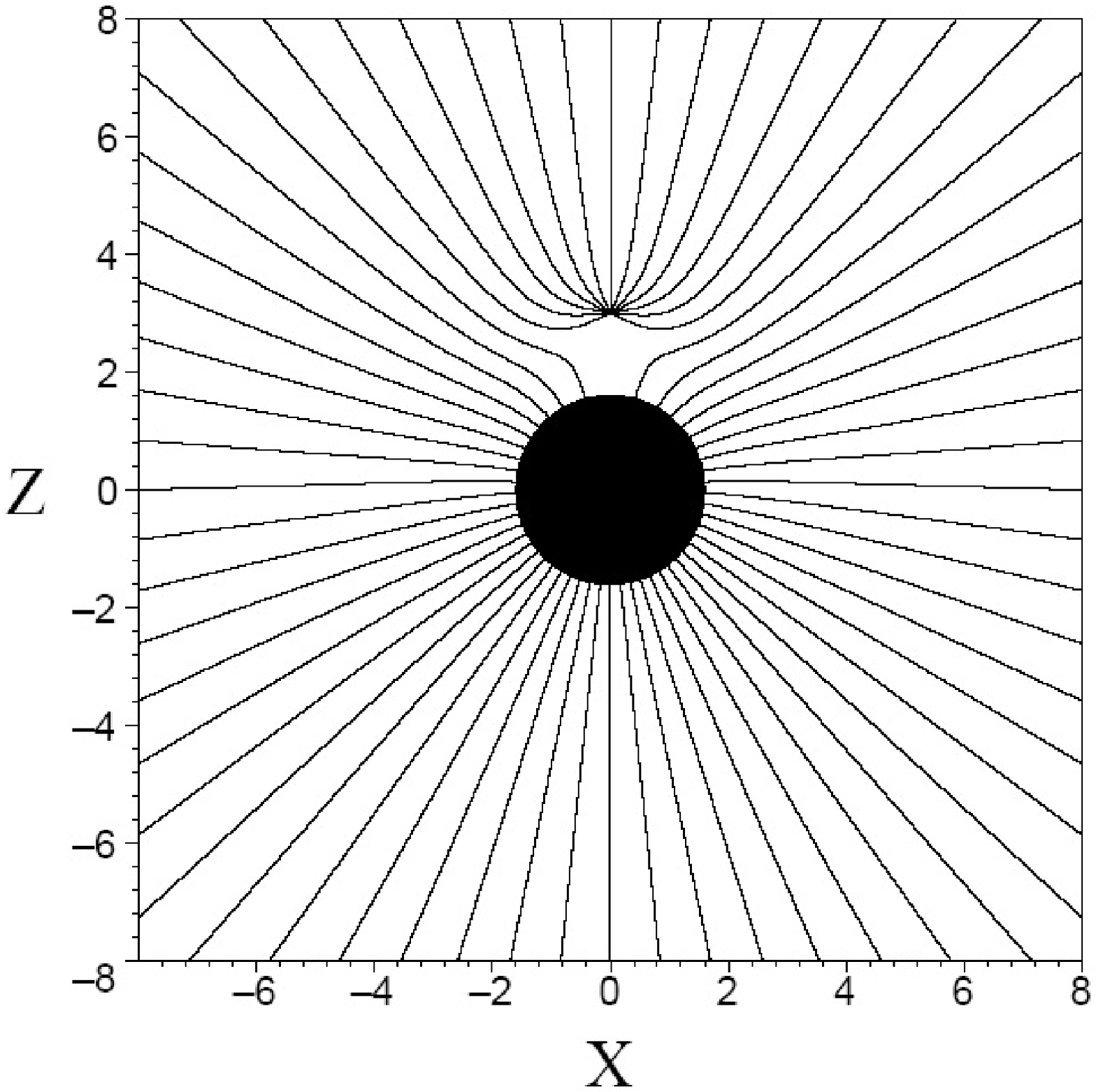}&
\includegraphics[scale=0.45]{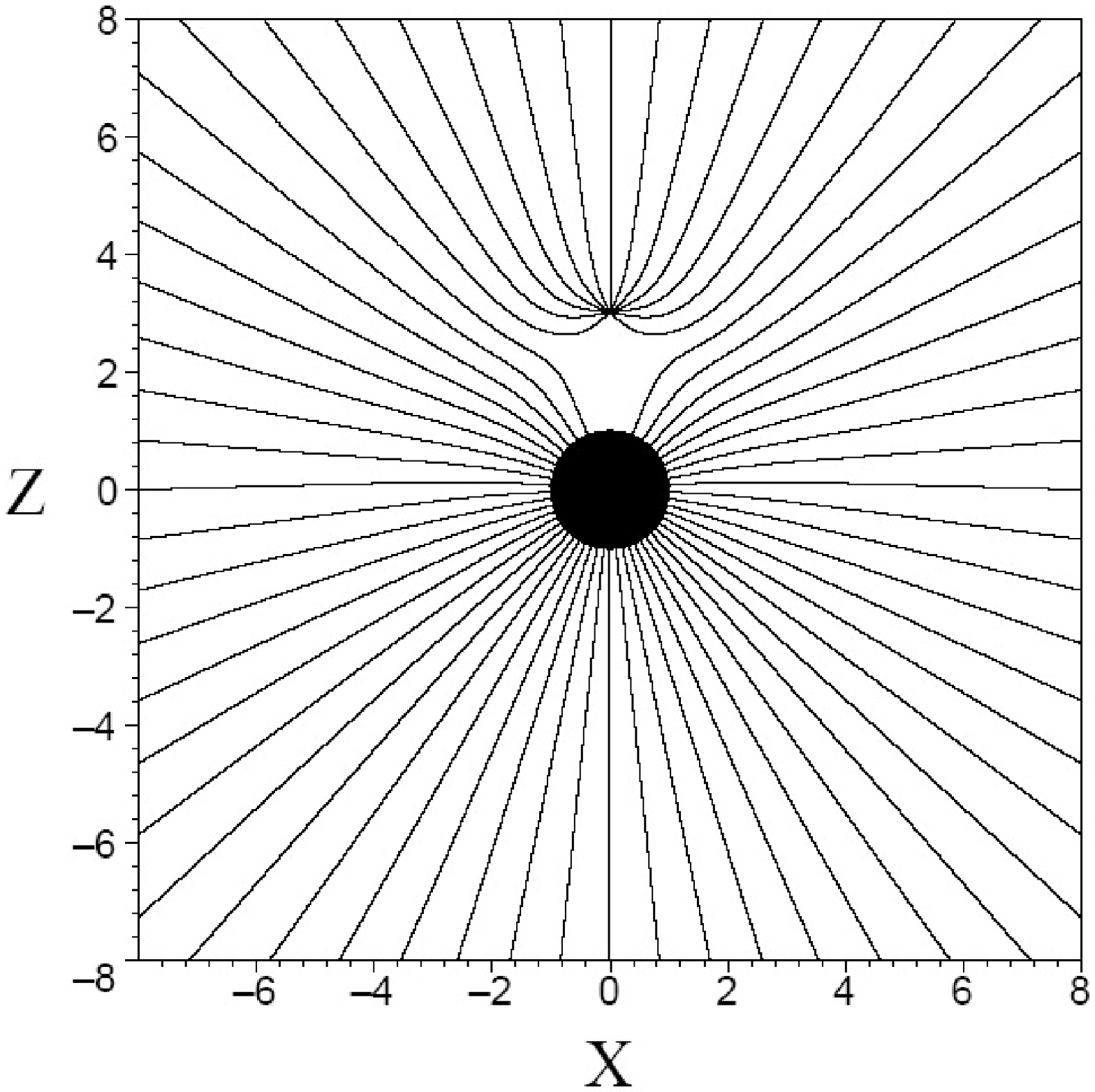}\\[0.4cm]
\mbox{(a)} & \mbox{(b)}\\
\end{array}$
\end{center}
\caption{Lines of force of the total electric field of the black hole and particle in the $X-Z$ plane ($X=r\sin \theta$, $Z=r\cos \theta$ are Cartesian-like coordinates) for charges of the same sign $q/Q=0.1$ and fixed parameter values  $b/\mathcal M=3$ and (a) ${Q}/{{\mathcal M}}=0.8$ and (b) ${Q}/{{\mathcal M}}=1$ respectively. 
The equilibrium condition (\ref{bonnoreqcond}) then implies that the ratio between the masses of the black hole and particle are given by (a) $m/\mathcal M\approx0.06$ and (b) $m/\mathcal M=0.1$ respectively.
The black hole horizon is located at (a) $r_+/{{\mathcal M}}=1.6$ and (b) $r_+/{{\mathcal M}}=1$ respectively.}
\label{fig:1}
\end{figure}

Fig.~\ref{fig:1} shows the lines of force of the total electric field of the whole system consisting of the black hole and particle for charges of the same sign both in the non-extreme (see Fig. (a)) and the extreme (see Fig. (b)) case.


We turn now to the study of the \lq\lq effective field'' of the perturbation induced by the massive charged particle on the background electric field.
We have to provide a satisfactory definition of such a field from both geometrical and physical point of view, separating the contribution due to the particle alone (which is first order) from the black hole one.
From the definition (\ref{pertelecompts}) of the perturbed electric field components we see that the contraction of the electromagnetic tensor with the observer four-velocity generates various first order terms which contribute to the electric field as well. In order to identify the field representing the net perturbation, i.e. with the black hole contribution being subtracted, we rather use the flux equation (\ref{gausspert3}). We require that the integration over a spherical $2$-surface $S$ centered at the origin gives the first order contribution $\Phi^{(1)}=4\pi q\vartheta(r-b)$ only to the total electric flux (\ref{gausspert}), i.e. the charge of the particle only (up to the $4\pi$ factor)
\begin{equation}
d\Phi^{(1)}=2\pi [{}^*{\tilde F}_{r\phi}^{(1)}dr +{}^*{\tilde F}_{\theta\phi}^{(1)}d\theta]\ .
\end{equation}
The \lq\lq effective field'' lines corresponding to the perturbation with the contribution of the black hole electric field being subtracted are thus defined as the lines of constant flux $d\Phi^{(1)}=0$, namely
\begin{equation}
0={}^*{\tilde F}_{r\phi}^{(1)}dr+{}^*{\tilde F}_{\theta\phi}^{(1)}d\theta\ .
\end{equation}

\begin{figure} 
\typeout{*** EPS figure 2}
\begin{center}
\includegraphics[scale=0.45]{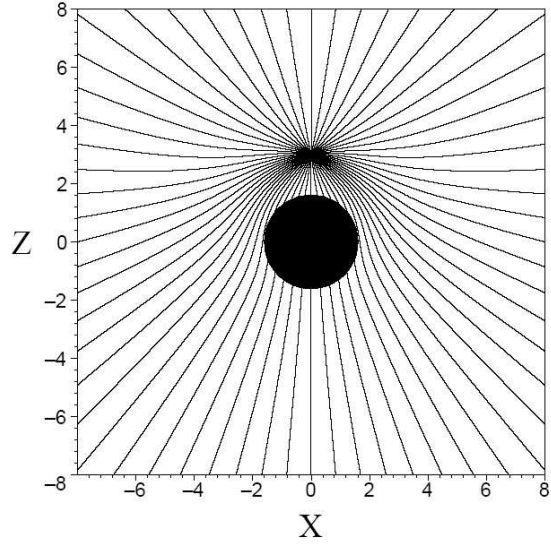}
\end{center}
\caption{Lines of force of the effective electric field of the sole particle in the non-extreme case for the same choice of parameters as in Fig.~{\ref{fig:1}} (a).
As explained in the text this \lq\lq effective field'' is obtained by subtracting the dominant contribution of the black hole own electric field to the total perturbed field, thus representing the net effect of the perturbation induced by the massive charged particle on the background field.}
\label{fig:2}
\end{figure}

In Fig.~{\ref{fig:2}} we show the behavior of the lines of force of the effective electric field of the sole particle in the non-extreme case for the same choice of parameters adopted in Fig.~{\ref{fig:1}} (a).

\subsection{The induced charge on the black hole horizon}

Following Hanni and Ruffini \cite{HR} we now compute the induced charge on the surface of the black hole horizon. 
Some lines of force intersect the horizon. If the particle is positively charged, at angles smaller than a certain critical angle the induced charge is negative and the lines of force cross the horizon. At angles greater than the critical angle the induced charge is positive and the lines of force extend out of the horizon. At the critical angle the induced charge density vanishes and the lines of force of the electric field are tangent to the horizon.
The total electric flux through the horizon and thus the total induced charge are zero.

The induced charge density on the horizon $\sigma^H(\theta)$ is defined in such a way that the amount of induced charge on an infinitesimal portion of the horizon sphere $r=r_+$ between $\theta=\theta_0$ and $\theta=\theta_0+d\theta$ equals $1/(4\pi)$ times the elementary flux across the same surface 
\begin{equation}
\frac1{4\pi}d\Phi\vert_{r_+}=\frac1{4\pi}2\pi {}^*{\tilde F}_{\theta\phi}^{(1)}\big\vert_{r_+}d\theta
=2\pi r_+^2\sigma^H(\theta)\sin\theta d\theta\ ,
\end{equation}
implying
\begin{equation}
\label{HRgauss}
\frac{{}^*{\tilde F}_{\theta\phi}^{(1)}\big\vert_{r_+}}{r_+^2\sin\theta} = 4\pi \sigma^H(\theta)\ .
\end{equation}
This can be identified with the surface version of the Gauss' law. 
The corresponding expression for the critical angle $\theta_{\rm (crit)}$ comes from the condition $\sigma^H(\theta_{\rm (crit)})=0$. 
Hence it results
\begin{eqnarray}
\label{sigmaRN}
\sigma^H(\theta)
&=&\frac{q}{4\pi r_+}\frac{\Gamma^2}{{\mathcal M}b-Q^2}\frac{\Gamma(1+\cos^2\theta)-2(b-\mathcal M)\cos\theta}{[b-\mathcal M-\Gamma\cos\theta]^2}\ , \\
\label{thetacritRN}
\theta_{\rm (crit)}
&=&\arccos\left[ \frac{b-\mathcal M-\sqrt{(b-\mathcal M)^2 -\Gamma^2}} {\Gamma} \right]\ .
\end{eqnarray}
Assuming then the black hole and particle both have positive charge, one can evaluate the total amount of negative charge induced on the horizon by the particle
\begin{eqnarray}
\label{Qindmeno}
Q_{\rm{ind}}^{(-)}&=&\int_\Sigma\sigma^H(\theta)d\Sigma
=2\pi r_+^2\int_0^{\theta_{\rm (crit)}}\sigma^H(\theta)\sin\theta d\theta
=-q\frac{\Gamma r_+}{{\mathcal M}b-Q^2}\cos\theta_{\rm (crit)}\ ,
\end{eqnarray}
where $d\Sigma=\sqrt{g_{\theta\theta} g_{\phi\phi}}\,d\theta\,d\phi$ and $\Sigma$ is the spherical 
cap $0 \leq \theta \leq \theta_{\rm (crit)}$.


The use of Gauss' law (\ref{HRgauss}) to define the induced charge density on the horizon requires some care, although it has been formulated in terms of the flux which is well defined everywhere including the horizon. 
Introduce an orthonormal frame adapted to the static observers (\ref{staticobs}) 
\begin{equation}
e_{\hat t}=U\ , \qquad e_{\hat r}=\frac1{\sqrt{\tilde g_{rr}}}\partial_r\ , \qquad e_{\hat \theta}=\frac1{\sqrt{\tilde g_{\theta\theta}}}\partial_\theta\ , \qquad e_{\hat \phi}=\frac1{\sqrt{\tilde g_{\phi\phi}}}\partial_\phi\ .
\end{equation}
Both $e_{\hat t}$ and $e_{\hat r}$ become null vectors on the horizon $r_+$. Therefore the observer is no more a physical observer and he cannot lead to the determination of the components of the electric fields. The frame itself is no more a physical frame; hence an alternative procedure is needed.

The first attempt to evaluate the electromagnetic fields in a well behaved frame at the horizon is due to Znajek \cite{znajek} and Damour \cite{damour}. Their main idea consists of surrounding the black hole by a special surface which is considered as the boundary of the black hole itself. Such a surface was identified with the horizon, since what happens inside the hole cannot affect at all the exterior region, and all the needed boundary conditions for physical quantities are imposed there. Znajek and Damour thus introduced a fictitious surface charge density as well as a surface current to terminate the electric as well as magnetic field components on the horizon.

Another possibility which has been pursued by Thorne and collaborators (see e.g. the monograph \cite{membrane}) consists of identifying the boundary of the black hole with a surface located slightly outside the event horizon, known as \lq\lq stretched horizon,'' endowed with surface density of charge and electric current.
It is timelike and is chosen so that it coincides with an equipotential surface of the lapse function $\alpha=\alpha_{SH}=\,\,$const ($\alpha_{SH}\ll1$) of a family of external reference observers with respect to which the fields are measured, typically the Zero Angular Momentum Observers (ZAMOs) for stationary axisymmetric spacetimes.
Their lapse function (which in our case is simply $\alpha=\sqrt{-\tilde g_{tt}}$) goes to zero at the horizon, leading to the pathological behavior of their adapted frame discussed above; in contrast, a freely falling observer crosses the horizon in a finite proper time and measures fields which are all finite there. 
From the transformation laws between electric and magnetic field quantities in these two frames it follows that the radial components as measured by ZAMOs remain finite, while the tangential ones blow up as $1/\alpha$ near the horizon. 
This divergence is eliminated by a regularization procedure consisting of multiplying the ill defined field components by the lapse function, so obtaining well behaved horizon fields \cite{TM}.
When evaluated at the stretched horizon these renormalized tangential fields are equal to the true-horizon fields defined by Znajek and Damour, to within fractional errors of order $\alpha_{SH}$, and are defined in such a way that in the limit $\alpha\to0$ they do not depend on the chosen position $\alpha_{SH}$ of the stretched horizon.
Obviously, does not exist any special membrane with physical properties outside the horizon of a black hole: the \lq\lq membrane paradigm'' is only an useful mathematical tool.

In our derivation, however, all this problematic has been by-passed using the flux equation which is well defined all the way to the horizon and has allowed to obtain the charge density following the same procedure introduced by Hanni and Ruffini.

\subsection{Embedding diagram}

The plots of Figs.~{\ref{fig:1}} and {\ref{fig:2}} actually shows a distorted view of the behavior of the electric field lines; we should rather look at their projection onto the corresponding embedding diagram, which gives the correct geometry allowing to visualize the spacetime curvature.  

The perturbed metric (\ref{belalekmetlin}) can be visualized as a 2-dimensional hyperboloid embedded in the usual euclidean 3-space with respect to the static observers (\ref{staticobs}) by suppressing the temporal and azimuthal dependence.
The induced metric of the constant time slice of the world sheet of the equatorial plane $\theta=\pi/2$ is given by
\begin{equation}
\label{embmetric}
^{(2)}d{\tilde s}^2=[1+{\bar {\mathcal H}}+k(r)]f(r)^{-1}dr^2+(1+{\bar {\mathcal H}})r^2d\phi^2\ ,
\end{equation}
where ${\bar {\mathcal H}}$ has to be evaluated at $\theta=\pi/2$.
For $r>r_+$ the coordinate $r$ is spacelike, so this metric can be embedded into the 3-dimensional euclidean space.
The flat-space line element written in cylindrical-like coordinates is given by
\begin{equation}
\label{3euclmetric}
^{(3)}ds^2=d\rho^2+\rho^2d\phi^2+dz^2\ .
\end{equation}
For the embedding surface in the parametric form $\rho=\rho(r)$, $z=z(r)$ the corresponding line element becomes
\begin{equation}
\label{2euclmetric}
^{(2)}ds^2=\gamma_{rr}dr^2+\rho^2d\phi^2\ ,
\end{equation}
where
\begin{equation}
\gamma_{rr}=\left(\frac{dz}{dr}\right)^2+\left(\frac{d\rho}{dr}\right)^2\ .
\end{equation}
Comparison with (\ref{embmetric}) implies
\begin{equation}
\gamma_{rr}=[1+{\bar {\mathcal H}}+k(r)]f(r)^{-1}\ , \qquad \rho=r\left(1+\frac{{\bar {\mathcal H}}}{2}\right)\ ,
\end{equation}
to first order of the perturbation.
The relation $\rho=\rho(r)$ is already given by the second equation and one can then numerically integrate the first equation, which is equivalent to 
\begin{equation}
\label{zdir}
\left(\frac{dz}{dr}\right)^2=[f(r)^{-1}-1][1+{\bar {\mathcal H}}+k(r)]+k(r)-r\partial_r{\bar {\mathcal H}}\ ,
\end{equation}
with the initial condition $z(r_+)=0$.

\begin{figure}
\typeout{*** EPS figure 3}
\begin{center}
\includegraphics[scale=0.45]{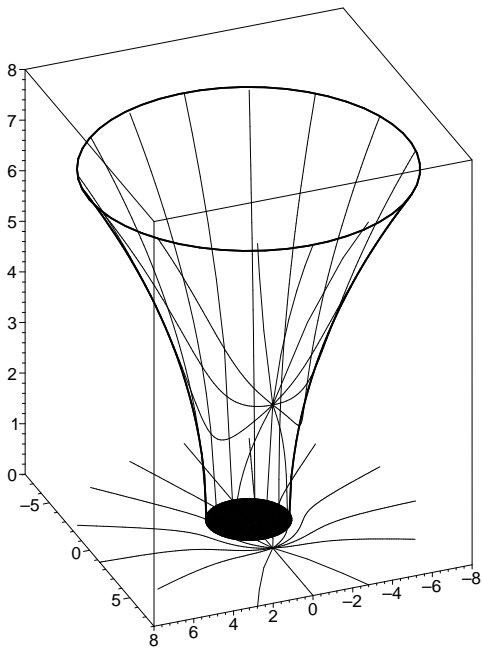}
\caption{The lines of force of the effective electric field of the sole particle are shown on an embedding diagram in the non-extreme case. The choice of the parameters is the same as in Fig.~{\ref{fig:2}}.
The projection of this embedding diagram onto the plane $z=0$ is also shown, and corresponds to the plot of Fig. \ref{fig:2} (the number of lines has been reduced for the sake of clarity).
Note that the coordinate $z$ defined by Eq. (\ref{zdir}) must not be confused with the Cartesian-like coordinate $Z$ of Figs. \ref{fig:1} and \ref{fig:2}.
It is interesting to compare and contrast this diagram obtained by direct integration of the embedding equations with the remarkable similarity of the qualitative picture drawn out of first principles by J. A. Wheeler (lower right hand side of Fig. 13, p. 58 of \cite{libro}), where the simplest problem of a charged test particle at rest near a Schwarzschild black hole was examined.}
\label{fig:3}
\end{center}
\end{figure}

Fig. \ref{fig:3} shows the embedding diagram of the lines of force of the effective electric field of the sole particle for the same choice of parameters as in Fig.~{\ref{fig:2}}. In the curved space the electric field is strictly radial at the horizon and the lines of force intersect the event horizon orthogonally.


\section{Extremely charged hole and the ``electric Meissner effect''}

\begin{figure} 
\typeout{*** EPS figure 2b}
\begin{center}
\includegraphics[scale=0.45]{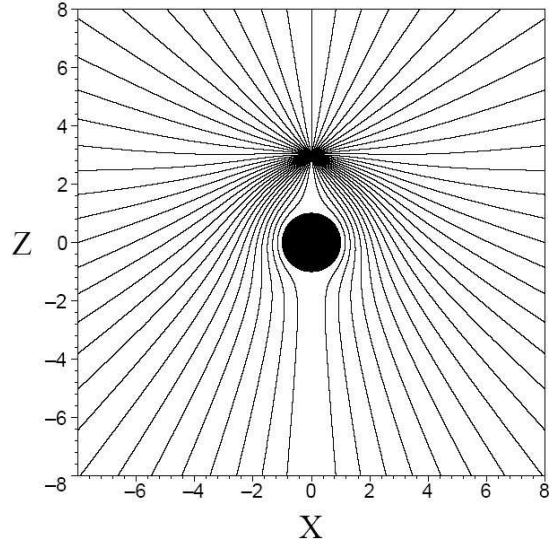}
\end{center}
\caption{Lines of force of the effective electric field of the particle alone in the extreme case for the same choice of parameters as in Fig.~{\ref{fig:1}} (b).
No lines of force intersect the black hole horizon in this case, leading to the the ``electric Meissner effect.''}
\label{fig:2b}
\end{figure}

So far the discussion has covered the case of a non-extreme black hole with $|Q|< {\mathcal M}$. 
Consider now the case $|Q|={\mathcal M}$ (implying $\Gamma=0$) of an extremely charged hole. 
Eq. (\ref{sigmaRN}) shows that the induced charge density on the horizon degenerates to zero for every value of the angle $\theta$; the critical angle (\ref{thetacritRN}) approaches the value $\pi/2$ and the amount of negative charge (\ref{Qindmeno}) induced on the horizon vanishes identically.
Therefore no lines of force cross the horizon, remaining tangent to it for every value of the polar angle, since every angle becomes critical: as the black hole approaches the extreme condition the electric field lines are thus pulled off the outer horizon and never intersect it when the black hole becomes extreme. 
The situation is summarized in Fig.~\ref{fig:2b}, where the behavior of the lines of force of the effective electric field of the sole particle in the extreme case is shown for the same choice of parameters as in Fig.~{\ref{fig:1}} (b).


It is worth noting that such an effect also occurs when treating the problem in the simplified approach of the test field approximation, whose hybrid nature has been already discussed in \cite{bgr_prd}. 
The components (\ref{fstarcompts}) of the dual of the electromagnetic tensor become in this case
\begin{equation}
{}^*{\tilde F}_{\theta\phi}^{\rm{test}}=-r^2\sin\theta\left[-\frac{Q}{r^2}+\partial_r V_{\rm{test}}\right]\ , \qquad
{}^*{\tilde F}_{r\phi}^{\rm{test}}=f(r)^{-1}\sin\theta\, \partial_\theta V_{\rm{test}}\ ;
\end{equation}
the corresponding induced charge density on the horizon $\sigma^H_{\rm{test}}(\theta)$ is then easily evaluated from Eq. (\ref{HRgauss}) and turns out to be related to Eq. (\ref{sigmaRN}) by
\begin{equation}
\label{sigmatest}
\sigma^H(\theta)=\frac{\Gamma b}{{\mathcal M}b-Q^2}\,\sigma^H_{\rm{test}}(\theta)\ .
\end{equation}
Eq. (\ref{sigmatest}) thus implies that the critical angle is the same as (\ref{thetacritRN}) and that the induced charge density on the horizon exhibits the same feature as above.
This is quite remarkable, since our result has been obtained within a more general framework, and both the gravitational and electromagnetic fields are different from those corresponding to the test field solution.
Therefore, we conclude that the electric Meissner effect is already present in the test field approximation, and it is confirmed at the perturbative level, taking into account all the feedback terms on both the background gravitational and electromagnetic fields due to the particle's mass and charge.

By using Eqs. (\ref{Qindmeno}) and (\ref{sigmatest}) the total amount of negative charge induced on the horizon by the particle in the case of test field is thus given by
\begin{eqnarray}
\label{Qindmenotest}
Q_{\rm{ind}\,\,\rm{test}}^{(-)}=\frac{{\mathcal M}b-Q^2}{\Gamma b}\,Q_{\rm{ind}}^{(-)}
=-q\frac{r_+}{b}\cos\theta_{\rm (crit)}\ ,
\end{eqnarray}
which is always greater than (\ref{Qindmeno}) for every fixed value of $|Q|/{\mathcal M}<1$, the multiplicative factor ranging from $1<({\mathcal M}b-Q^2)/(\Gamma b)<{\mathcal M}/\Gamma$ for $r_+<b<\infty$ and correspondingly $\infty>q/m>{\mathcal M}/Q$, as from Eq. (\ref{bonnoreqcond}).
In the extreme case $|Q|/{\mathcal M}=1$ the induced charge vanishes in both cases.

The ``electric Meissner effect'' above described is suitable to a conceptual interpretation in terms of the nature of the Reissner-Nordstr\"om solution.
When the black hole is not extreme the particle induces charge on the horizon, and accordingly the electric field lines terminate on it. When the black hole becomes extreme no further charge induction is possible without turning the black hole into a naked singularity, and coherently the electric field lines no more cross the horizon.   
Therefore, in a sense, the black hole \lq\lq rejects to turn itself into a naked singularity.''

\section{Conclusions}

The properties of the electric field of the charged two-body system consisting of a Reissner-Nordstr\"om black hole and a charged massive particle at rest, recently obtained in the framework of first order perturbation theory adopting the Regge-Wheeler gauge, have been further investigated.  
The analysis of the field lines carried out using the definition of the constant flux surfaces has shown the existence of the ``electric Meissner effect,'' namely the phenomenon of expulsion of the lines of force from the outer horizon as the hole becomes extreme, once the dominating background contribution due to the hole own electric field is properly subtracted.
This analysis thus extends to the electric field analogous results found by Bi{\v c}\'ak and Dvo{\v r}\'ak \cite{bicdev2} concerning magnetic fields. They showed that the Reissner-Nordstr\"om black hole behaves like a ``superconductor'' as approaching the extreme condition in the presence of a magnetic field, in close similarity with the ordinary magnetic Meissner effect. 
The electric Meissner effect has instead introduced a novel concept, since it has no classical analogue and is a pure general relativistic effect. 

The electric Meissner effect is a feature already present within the test field approximation, i.e. by neglecting the backreaction on the background metric and electromagnetic field due to the particle's mass and charge, as already pointed out by Ruffini \cite{elbaproc}.
This result by itself has justified addressing this problem in the more general Zerilli approach duly taking into account all the first order perturbations. 
Our solution thus provides a useful tool to explore the limits of validity of the test field approach, which has been largely used in the current literature.

Recently Belinski and Alekseev \cite{belinski} obtained an exact two-body solution to the Einstein-Maxwell equations for a Reissner-Nordstr\"om black hole in equilibrium with a naked singularity. 
We have shown in the Appendix that the Belinski-Alekseev solution, once linearized with respect to the mass and charge of the naked singularity, coincides with our solution.
In this limit also their equilibrium condition exactly reduces to our equilibrium condition. 
Our first order result therefore confirms the validity and offers a tool for the physical interpretation of the Belinski-Alekseev solution, taking advantage from the unambiguity in the definition of masses and charges as well as of the distance between the bodies due to the linearity of the interaction.

Our solution thus represents a \lq\lq bridge'' between the test field solution, which neglects all the feedback terms, and the exact two-body solution, which takes into account all the non-linearity of the interaction.

It is well known that, classically, a two-body system of massive charged particles with masses $M_1$ and $M_2$ and charges $Q_1$ and $Q_2$ can be in equilibrium for any value of their radial distance if $Q_1Q_2=M_1M_2$. 
This result has been proved to hold also in the fully relativistic case if and only if the objects have $Q_1=M_1$ and $Q_2=M_2$, i.e. are both extremely charged black holes \cite{maj,pap}.
Actually an equilibrium exists for any number of collinear extreme bodies, and the solution can be further extended to the case of angular momentum \cite{ruff}.
The non-extreme case is very different: the classical result can {\it never} be obtained even in the perturbative regime. There is always only one equilibrium configuration allowed, which has no classical counterpart.
The nature of the equilibrium has to be further studied and has certainly to do with the extended field description of the black hole and the naked singularity, which is described in the general relativistic perturbation approach by a convergent series of infinite multipoles expressible in a closed analytic form.

\begin{acknowledgments}
The authors are indebted to Profs. J. Ehlers, T. Damour and  R. T. Jantzen for a critical reading of the manuscript.
\end{acknowledgments}

\appendix

\section{The linearized form of the Belinski-Alekseev solution}

The Belinski-Alekseev solution in Weyl cylindrical coordinates $(t,\rho,z,\phi)$ is given by
\begin{equation}
\label{belalekmetric}
ds^2=-Hdt^2+f[d\rho^2+dz^2]+\frac{\rho^2}{H}d\phi^2\ ,
\end{equation}
with electromagnetic 4-potential 
\begin{equation}
\label{belalekpot}
A=-\Phi dt\ ,
\end{equation}
using the notation of \cite{belinski}, but maintaining our signature conventions; the functions $H$, $f$ and $\Phi$ depend on coordinates $\rho$ and $z$ only.
It is useful to introduce bipolar coordinates consisting of two pairs of spheroidal coordinates $(r_1,\theta_1)$ and $(r_2,\theta_2)$, in terms of which the metric functions take their most simple form
\begin{eqnarray}
\label{metricfuncts}
H&=&\mathcal{D}^{-2}[(r_1-m_1)^2-\sigma_1^2+\gamma^2\sin^2\theta_2][(r_2-m_2)^2-\sigma_2^2+\gamma^2\sin^2\theta_1]\ , \nonumber\\
f&=&\mathcal{D}^2[(r_1-m_1)^2-\sigma_1^2\cos^2\theta_1]^{-1}[(r_2-m_2)^2-\sigma_2^2\cos^2\theta_2]^{-1}\ , \nonumber\\
\Phi&=&\mathcal{D}^{-1}[(e_1-\gamma)(r_2-m_2)+(e_2+\gamma)(r_1-m_1)+\gamma(m_1\cos\theta_1+m_2\cos\theta_2)]\ , 
\end{eqnarray}
where
\begin{equation}
\mathcal{D}=r_1r_2-(e_1-\gamma-\gamma\cos\theta_2)(e_2+\gamma-\gamma\cos\theta_1)\ .
\end{equation}
The relations between bipolar coordinates and Weyl coordinates are given by
\begin{eqnarray}
\label{bipolar}
&&\rho=\sqrt{(r_1-m_1)^2-\sigma_1^2}\sin\theta_1\ , \qquad z=z_1+(r_1-m_1)\cos\theta_1\ , \nonumber\\
&&\rho=\sqrt{(r_2-m_2)^2-\sigma_2^2}\sin\theta_2\ , \qquad z=z_2+(r_2-m_2)\cos\theta_2\ .
\end{eqnarray}
The quantities $m_1$, $m_2$ and $e_1$, $e_2$ represent interacting masses and physical charges of the two Reissner-Nordstr\"om sources, which are located along the symmetry axis at the points $z=z_1$ and $z=z_2$ respectively. 
The parameters $\gamma$, $\sigma_1$ and $\sigma_2$ are determined by the relations
\begin{equation}
\label{constex}
\sigma_1^2=m_1^2-e_1^2+2e_1\gamma\ , \qquad
\sigma_2^2=m_2^2-e_2^2-2e_2\gamma\ , \qquad
\gamma=\frac{m_2e_1-m_1e_2}{\ell+m_1+m_2}\ , 
\end{equation}
with $\ell=z_2-z_1$ characterizing the distance between the sources.
The functions (\ref{metricfuncts}) satisfy the system of Einstein-Maxwell equations if the following equilibrium condition holds
\begin{equation}
\label{equil}
m_1m_2=(e_1-\gamma)(e_2+\gamma)\ .
\end{equation}

The solution (\ref{belalekmetric})--(\ref{belalekpot}) thus describes the field of a Reissner-Nordstr\"om black hole in equilibrium with a naked singularity whose mass, charge and distance parameters must satisfy the equilibrium condition (\ref{equil}).
Let the black hole be located about the origin $z_1=0$, whereas the naked singularity be at the point $z_2=\ell$.
Before proceeding with the linearization of the solution it is convenient to express all functions in terms of a single coordinate patch, inverting the relations (\ref{bipolar})
\begin{equation}
\label{trasfinv}
r_{1,2}=m_{1,2}+\frac{1}{2}(R_{1,2}^{+}+R_{1,2}^{-})\ , \qquad \theta_{1,2}=\arctan\left[\frac{\rho}{z}\frac{R_{1,2}^{+}+R_{1,2}^{-}}{\sqrt{(R_{1,2}^{+}+R_{1,2}^{-})^2-4\sigma_{1,2}^2}}\right]\ ,
\end{equation}
where
\begin{equation}
R_1^{\pm}=\sqrt{\rho^2+(z\pm \sigma_1)^2}\ , \qquad
R_2^{\pm}=\sqrt{\rho^2+(z-\ell\pm \sigma_2)^2}\ .
\end{equation}
Assume that the mass $m_2$ and charge $e_2$ of the naked singularity are small if compared with black hole mass $m_1$ and charge $e_1$, the ratio $e_2/m_2$ remaining finite. The equilibrium condition (\ref{equil}) thus gives 
\begin{equation}
\label{eqcondtest}
(\ell+m_2)(m_1m_2-e_1e_2)=(m_1e_2-m_2e_1)e_2\ ,
\end{equation}
to first order.
Belinski and Alekseev showed that this relation reproduces exactly our equilibrium condition (\ref{bonnoreqcond}) if the rest mass parameter $\mu_2$ is introduced instead of the mass $m_2$ as smallness parameter in the linearization procedure defined by the relation
\begin{equation}
m_2=\mu_2\sqrt{1-\frac{2m_1}{\ell+m_1}+\frac{e_1^2}{(\ell+m_1)^2}}+\frac{e_1e_2}{\ell+m_1}\ ;
\end{equation}
Eq. (\ref{eqcondtest}) thus takes the form
\begin{equation}
m_1-\frac{e_1^2}{\ell+m_1}=\frac{e_1e_2}{\mu_2}\sqrt{1-\frac{2m_1}{\ell+m_1}+\frac{e_1^2}{(\ell+m_1)^2}}\ ,
\end{equation}
or equivalently
\begin{equation}
\frac{e_2}{\mu_2}=\frac{\ell m_1+\Gamma_1^2}{e_1L}\ , \qquad
L^2=\ell^2-\Gamma_1^2\ , \qquad
\Gamma_1^2=m_1^2-e_1^2\ ,
\end{equation}
which coincides with Eq. (\ref{bonnoreqcond}) with the identification $\ell=b - m_1$, $m_1=\mathcal{M}$, $e_1=Q$, $\mu_2=m$, $e_2=q$.

The linearized expressions for the constants (\ref{constex}) turn out to be
\begin{equation}
\sigma_1=\Gamma_1-\mu_2\frac{\Gamma_1}{L}\ , \qquad
\sigma_2=i\mu_2\frac{\Gamma_1}{e_1}\ , \qquad
\gamma=\mu_2\frac{\Gamma_1^2}{e_1L}\ . 
\end{equation}
Substitute these quantities into the solution (\ref{metricfuncts}), after replacing the bipolar coordinates with Weyl coordinates using the transformations (\ref{trasfinv}); retaining terms up to the linear order in $\mu_2$ the metric functions and electrostatic potential then take the form
\begin{equation}
H=H_0+\mu_2 H_1\ , \qquad 
f=f_0+\mu_2 f_1\ , \qquad 
\Phi=\Phi_0+\mu_2 \Phi_1\ ,
\end{equation}
where 
\begin{eqnarray}
\label{RNsolw}
H_0&=&\frac{(S_1^{+}+S_1^{-})^2-4\Gamma_1^2}{(2m_1+S_1^{+}+S_1^{-})^2}\ , \qquad 
f_0=\frac{(2m_1+S_1^{+}+S_1^{-})^2}{4S_1^{+}S_1^{-}}\ , \nonumber\\
\Phi_0&=&\frac{2e_1}{2m_1+S_1^{+}+S_1^{-}}\ , \qquad 
S_1^{\pm}=\sqrt{\rho^2+(z\pm \Gamma_1)^2}\ ,
\end{eqnarray}
are the corresponding background functions. 
After long calculations the first order terms turn out to be given by
\begin{eqnarray}
\label{pertsolw}
H_1&=&H_0\left\{\frac{\Phi_0}{2Le_1S_1^{+}S_1^{-}}[m_1(S_1^{+}-S_1^{-})^2+2\Gamma_1^2(S_1^{+}+S_1^{-})]-{\bar H}\right\}\ , \nonumber\\
f_1&=&f_0\left\{(1-2H_0f_0)\frac{(S_1^{+}-S_1^{-})^2}{2LS_1^{+}S_1^{-}}+2\frac{m_1}{e_1}\frac{\Phi_0}{L}(1-H_0f_0)+{\bar H}\right\}\ , \nonumber\\
\Phi_1&=&\Phi_0\left\{\frac{\bar H}{4e_1^2}[m_1(S_1^{+}+S_1^{-})+2\Gamma_1^2]+\frac{\Gamma_1^2}{Le_1^2}-\frac{S_1^{+}+S_1^{-}}{4LS_1^{+}S_1^{-}}\frac{(S_1^{+}-S_1^{-})^2-4\Gamma_1^2}{2m_1+S_1^{+}+S_1^{-}}\right\}\ , 
\end{eqnarray}
where 
\begin{equation}
\bar H=\frac{2}{L}\frac{\ell(S_1^{+}+S_1^{-})-\Gamma_1(S_1^{+}-S_1^{-})}{S_2(2m_1+S_1^{+}+S_1^{-})}\ , \qquad 
S_2=\sqrt{\rho^2+(z-\ell)^2}\ .
\end{equation}
In order to compare this solution with our solution it is convenient to express both the perturbed metric and electromagnetic field in terms of Schwarzschild-like coordinates centered about the black hole using the first transformation of (\ref{bipolar}), whose linearized form is given by
\begin{equation}
\label{bipolarlin}
\rho=\sqrt{(r_1-m_1)^2-\Gamma_1^2}\sin\theta_1+\mu_2\frac{\Gamma_1^2}{L}\frac{\sin\theta_1}{\sqrt{(r_1-m_1)^2-\Gamma_1^2}}\ , 
\qquad z=(r_1-m_1)\cos\theta_1\ .
\end{equation} 
The linearized form of the metric (\ref{belalekmetric}) thus becomes exactly as in Eq. (\ref{belalekmetlin}) with the identifications $\ell=b - m_1$, $m_1=\mathcal{M}$, $e_1=Q$, $\mu_2=m$, $e_2=q$ and by suppressing the subscript $_1$ on the coordinates $r_1$ and $\theta_1$.
The linearized electromagnetic field is the same as in Eqs. (\ref{RNemfieldpertnonextr}).


\begin{thebibliography}{00}

\bibitem{bgr_pla}
D. Bini, A. Geralico, and R. Ruffini, 
Phys.\ Lett.\ A {\bf 360}, 515 (2007).

\bibitem{bgr_prd}
D. Bini, A. Geralico, and R. Ruffini, 
Phys.\ Rev.\ D {\bf 75}, 044012 (2007).

\bibitem{Zerilli}
F. J. Zerilli, 
Phys.\ Rev.\ D {\bf 9}, 860 (1974).

\bibitem{ReggeW}
T. Regge and J. A. Wheeler, 
Phys.\ Rev. {\bf 108}, 1063 (1957).

\bibitem{jrz2}
M. Johnston, R. Ruffini, and F. J. Zerilli, 
Phys.\ Lett.\ B {\bf 49}, 185 (1974).

\bibitem{jrz1}
M. Johnston, R. Ruffini, and F. J. Zerilli, 
Phys.\ Rev.\ Lett. {\bf 31}, 1317 (1973).

\bibitem{HR}
R. Hanni and R. Ruffini, 
Phys.\ Rev.\ D {\bf 8}, 3259 (1973).

\bibitem{leaute}
B. Leaute and B. Linet, 
Phys.\ Lett.\ A {\bf 58}, 5 (1976).

\bibitem{bonnor}
W. B. Bonnor, 
Class.\ Quant.\ Grav. {\bf 10}, 2077 (1993).

\bibitem{bicdev2}
J. Bi{\v c}\'ak and L. Dvo{\v r}\'ak, 
Phys.\ Rev.\ D {\bf 22}, 2933 (1980).

\bibitem{moncrief1}
V. Moncrief,
Phys.\ Rev.\ D {\bf 9}, 2707 (1974). 

\bibitem{moncrief2}
V. Moncrief,
Phys.\ Rev.\ D {\bf 10}, 1057 (1974). 

\bibitem{moncrief3}
V. Moncrief,
Phys.\ Rev.\ D {\bf 12}, 1526 (1975). 

\bibitem{bicak}
J. Bi{\v c}\'ak,  
Czech.\ J.\ Phys.\ B {\bf 29}, 945 (1979).

\bibitem{bicledv}
J. Bi{\v c}\'ak and T. Ledvinka, 
Il Nuovo\ Cimento\ B {\bf 115}, 739 (2000).

\bibitem{emparan}
A. Chamblin, R. Emparan, and G. W. Gibbons,  
Phys.\ Rev.\ D {\bf 58}, 084009 (1998).

\bibitem{belinski}
G. A. Alekseev and V. A. Belinski, 
Phys.\ Rev.\ D {\bf 76}, 021501(R) (2007). 

\bibitem{alek}
G. A. Alekseev,
Sov.\ Phys.\ Dokl.\ {\bf 30}, 565 (1985);
Proc. Steklov Inst. Math. {\bf 3}, 215 (1988).

\bibitem{belalekgrqc}
V. A. Belinski and G. A. Alekseev, 
arXiv: gr-qc/0710.2515.  

\bibitem{manko2}
V. S. Manko, 
arXiv: gr-qc/0710.2158.

\bibitem{manko}
N. Bret\'on, V. S. Manko, and J. Aguilar-S\'anchez, 
Class.\ Quantum\ Grav. {\bf 15}, 3071 (1998).

\bibitem{sibgatullin}
N. R. Sibgatullin, 
{\it Oscillations and Waves in Strong Gravitational and Electromagnetic Fields}
(Nauka, Moscow, 1984) [English translation (Springer-Verlag, Berlin, 1991)].

\bibitem{maj}
S. M. Majumdar, 
Phys.\ Rev.\ {\bf 72}, 390 (1947).

\bibitem{pap}
A. Papapetrou, 
Proc.\ R.\ Irish\ Acad.\ {\bf 51}, 191 (1947).

\bibitem{bgj}
D. Bini, C. Germani, and R. T. Jantzen, 
Int.\ J.\ Mod.\ Phys.\ D {\bf 10}, 633 (2001). 

\bibitem{znajek}
R. L. Znajek, 
Mon.\ Not.\ R.\ Astron.\ Soc. {\bf 185}, 833 (1978).

\bibitem{damour}
T. Damour, 
Phys.\ Rev.\ D {\bf 18}, 3598 (1978).

\bibitem{membrane}
K. S. Thorne, R. H. Price, and D. A. Macdonald,
\textit{Black Holes: the Membrane Paradigm} (Yale University Press, New Haven, 1986).

\bibitem{TM}
K. S. Thorne and D. A. Macdonald,
Mon.\ Not.\ R.\ Astron.\ Soc. {\bf 198}, 339 (1982).

\bibitem{libro}
M. Rees, R. Ruffini, and J. A. Wheeler, 
{\it Black Holes, Gravitational Waves and Cosmology}
(Gordon and Breach, New York, 1973).

\bibitem{elbaproc}
R. Ruffini, 
Il Nuovo\ Cimento B {\bf 119}, 785 (2004).

\bibitem{ruff}
L. Parker, R. Ruffini, and D. Wilkins,
Phys.\ Rev.\ D {\bf 7}, 2874 (1973).



\end{thebibliography}
\end{document}